\documentclass[fp,twocolumn]{jpsj3}
\usepackage{amsmath,amsbsy,amssymb}
\usepackage{graphicx}
\usepackage{amsmath}
\usepackage{braket}
\usepackage{bm}
\usepackage{color}
\usepackage{txfonts}
\usepackage{cite,overcite}

\usepackage{color,xcolor}
\usepackage[normalem]{ulem}

\renewcommand{\figurename}{Fig.}

\begin{document}

\title{Hybrid Higher-Order Skin Topological Modes in the Two-Dimensional Su-Schrieffer-Heeger Model with Nonreciprocal Hoppings}
\author{Hiromasa Wakao}
\date{\today}

\inst{%
Graduate School of Pure and Applied Sciences, University of Tsukuba, Tsukuba, Ibaraki 305-8571, Japan
}%

\abst{
The coexistence of edge states and skin effects
provides the topologically protected localized states at the corners of two-dimensional systems.
In this paper,
we realize
such corner states
in the two-dimensional Su-Schrieffer-Heeger model with the nonreciprocal hoppings.
For the characterization of the real line gap topology,
we introduce the $\mathbb{Z}_4$ Berry phase protected by generalized four-fold rotational symmetry.
From the physical picture of the adiabatic connection,
we find that the value of the $\mathbb{Z}_4$ Berry phase predicts the position of edge states.
Additionally,
by using the winding number,
we characterize the point gap topology of the edge spectra.
From the results of these characterizations by the first-order topological invariants,
we find that the pair of values of the $\mathbb{Z}_4$ Berry phase and the winding number yields
the position of the topologically protected localized states.
}

\maketitle

\section{Introduction \label{sec:introduction-}}
Topological materials which are characterized by the topology in the bulk have been actively pursued~\cite{Hatsugai19930929,Hatsugai19931015,Kane20050928,Teo20080723,Qi20080924,Hasan20100908,Qi20111014,Sato20160328}.
Additionally, the description of these materials are enriched by symmetry~\cite{Fu20060908,Fu20070307,Moore20070326,Zhang20090510,Roy20090521,Hsieh20090928,Chiu20160831}.
For instance, the presence of time-reversal symmetry provides $\mathbb{Z}_2$ topological properties~\cite{Fu20060908}.
Moreover,
it turns out that crystalline symmetry
plays an key role for the determination of the band topology~\cite{Zak19890605,Fu20110308,Hatsugai20110630,Fang20120910,Slager20131216,Kim20150717,Po20170730,Kruthoff20171222,Ono20180925,Kariyado20180615,Watanabe20181228}.
For example,
the coexistence of the time-reversal symmetry and inversion symmetry
enables us to introduce the form of the $\mathbb{Z}_2$ indicator
determined from the parity at the high-symmetry points in the two-dimensional Brillouin zone~\cite{Fu20110308}.

In the past decades,
the notion of the topological protection has been extended to a new area:
the non-Hermitian systems~\cite{Hatano19960715,Bender19980615,Rotter20090320,Sato20120701,Ashida20160916,Ashida20170608}.
Since the eigenvalues may become complex, many exotic topological phenomena have been provided~\cite{Hu20111007,Esaki20111117,Liang20130117,Gong20170518,Shen20180406,Yao20180924,Gong20180924,Carlstrom20181015,Budich20190124,Carlstrom20190424,Okuma20190829,Yoshida20190915,Kawabata20191021,Longhi20230213,Okuma202303},
such as the skin effects~\cite{Kunst20180711,Yao20180821,Lee20190508,Borgina20200207,Zhang20200915,Okuma20200225,Okuma20210428,Sun20210803,Lin20230703,Yoshida20230025} and
exceptional points (rings)~\cite{Zhen20150909,Lee20160401,Leykam20170123,Kozzi20170819,Alvarez20180307,Yoshida20180726,Kawabata20190809,Wojcik20200515,Bergholtz20210224,Yang20210225,Rui20220601}.
Particularly,
a distinct difference from the case in Hermitian systems
is the emergence of two types of the gap, i.e. a line gap and a point gap~\cite{Kawabata20191021}.
While the line gap topology can be understood by mapping onto the Hermitian cases,
the nontrivial point gap topology provides
novel phenomena without analogies in the Hermitian case, such as
the anomalous sensitivity of the energy spectrum to the presence or absence of boundaries, i.e., the skin effect.
In addition, as is the case for the Hermitian systems, these unique topological phenomena are enriched by symmetry~\cite{Okugawa20190124,Zhou20190211,Liu20190304,Yoshida20190305,Zhou20190606,Yoshida20200617,Okugawa20210521,Vecsei20210521,Shiozaki20210719,Mandal20211025,Delplace20211025,Chen20220224,Tsubota20220523,Yoshida20220207,Wakao20220613,Halder20221229,Tanaka20230615}.
These phenomena have been reported in a wide range of systems such as
open quantum systems~\cite{Lee20160401,Xu20170127,Yoshida20200917}, electrical circuits~\cite{Ezawa20190521,Hofmann20190621,Ezawa20190802,Yoshida20200617,
Helbig20200601,Hofmann20200602}, phononic systems~\cite{Yoshida20190822,Ghatak20200909,Scheibner20200909}, and photonic crystals~\cite{Guo20090827,Ruter20100124,Feng20170930,Isobe20210907}, and so on.

Along with the above developments,
a novel class of topological insulators has been extensively investigated, which is called higher-order topological insulators.
In $d$-dimensional systems, while the conventional topological insulators host $(d-1)$-dimensional boundary states,
higher-order topological insulators host $(d-2)$- or fewer dimensional boundary states~\cite{Benalcazar20170707,Benalcazar20171211,Ezawa20180112,Schindler20180101,Kunst20180611,Schindler20180730,Hayashi20180821,Okugawa20191204,Song20171211,Khakaf20180525,Matsugatani20180916,Trifunovic20190122,Benalcazar20190626,Kudo20190907,Mizoguchi20190805,Araki20200109,Takahashi20200312,Wakao20200330,Bunney20220110}.
For the characterization of the topological phase and the prediction of the emergence of these unique boundary states, many efforts have been devoted.
Remarkably, the presence of crystalline symmetry plays an important role for the description of the higher-order topology~\cite{Benalcazar20170707,Benalcazar20171211,Ezawa20180112,Schindler20180101,Kunst20180611,Song20171211,Khakaf20180525,Matsugatani20180916,Trifunovic20190122,Benalcazar20190626,Kudo20190907,Mizoguchi20190805,Araki20200109,Takahashi20200312,Wakao20200330,Bunney20220110}.

Recently,
the notion of the higher-order topology has also been extended to non-Hermitian systems~\cite{Liu20190220,Edvardsson20190227,Luo20190813,Yu20210105}.
In particular,
the coexistence of edge states and the skin effect in two-dimensional systems provides the topologically protected corner states,
which are called the hybrid higher-order skin topological modes~\cite{Lee20190702,Zhang20210901,Zou20211010,Li20220601,Zhu20220727,Zhu20230712,Sun20230809}.
These novel states in non-Hermitian systems represent a kind of the higher-order skin effect~\cite{Kawabata20200917,Okugawa20201216,Fu202120210119}.
Furthermore,
these unique localized states are characterized by the first- and second-order topological invariants
such as the Chern number~\cite{Lee20190702,Li20220601}, the Wess-Zumino term~\cite{Kawabata20200917}, and the symmetry indicator~\cite{Okugawa20201216,Shiozaki20210719}.

In this paper,
we demonstrate the realization of the hybrid higher-order skin topological modes in the two-dimensional Su-Schrieffer-Heeger (SSH) model with the nonreciprocal hoppings.
Specifically, we characterize the first-order topology by using the $\mathbb{Z}_4$ Berry phase protected by generalized four-fold rotational symmetry and the winding number on the cylinder geometry.
From the nontrivial value of the $\mathbb{Z}_4$ Berry phase, we show that there exists the bulk-edge correspondence between this topological invariant and edge states.
Additionally, by using the winding number, we realize the skin effect arising from the energy spectra of edge states.
Combining the $\mathbb{Z}_4$ Berry phase and the nontrivial point gap topology of these edge spectra, we demonstrate that the pair of these topological invariants indicates the position of the localized corner in two dimension.

\begin{figure*}[t]
 \centering
  \centering
\includegraphics[trim={0cm 0cm 0cm 0cm},width =0.9\hsize]{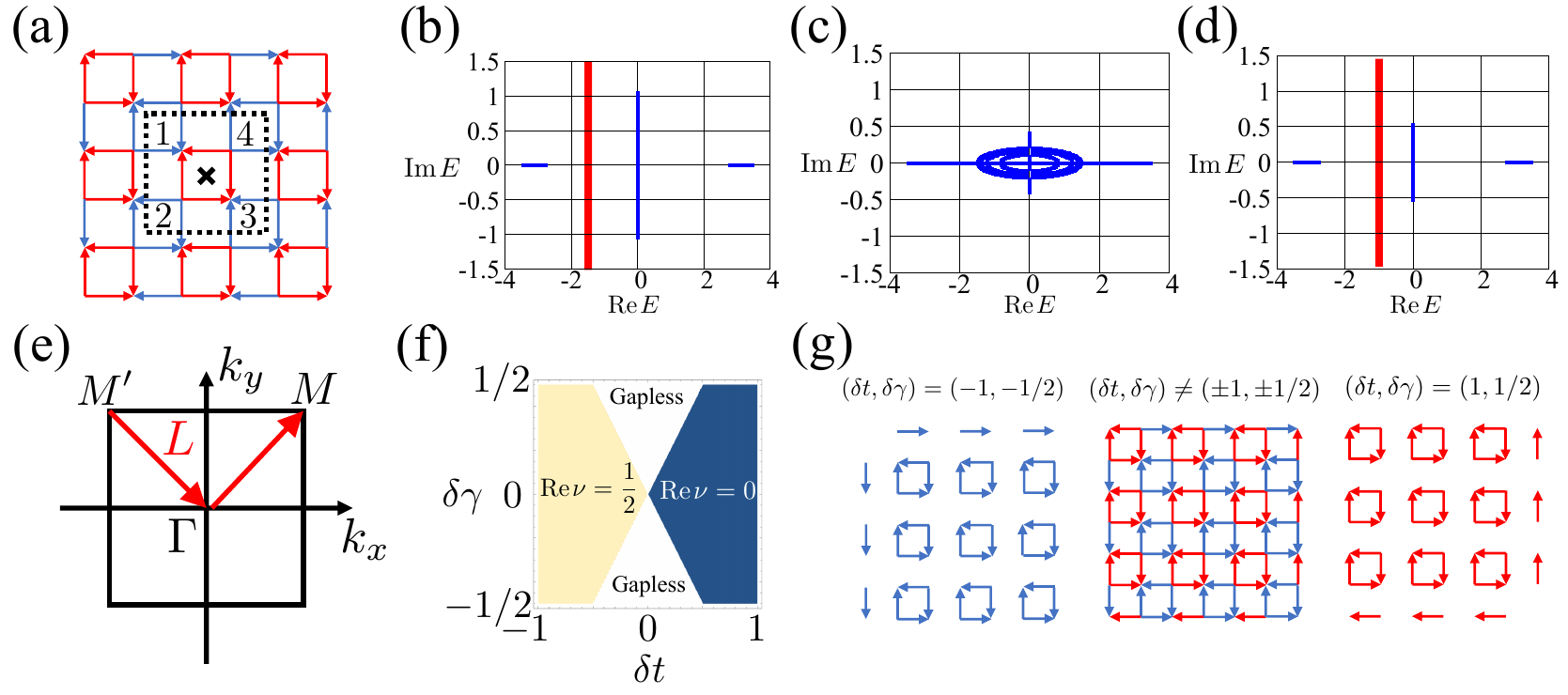}
 \caption{(Color online)
 (a) Schematic figure of the two-dimensional SSH model with the nonreciprocal hoppings.
 The unit cell is enclosed by the black dashed box.
 The cross denotes the center of the generalized four-fold rotation.
 (b), (c), and (d) show the energy spectrum for $(\delta t , \delta \gamma) = (-0.7,0.1)$, $(0,0.1)$ and $(0.7,0.1)$.
 Red lines denote the real line gaps where we focus.
 (e) Sketch of the Brillouin zone and the path of the integration in Eq.~(\ref{eq:5}).
 The path $L$ connects high-symmetry points $\Gamma$ and $M$ ($M^{\prime}$).
 (f) Color plot of the $\mathbb{Z}_4$ Berry phase.
 (g) Schematic figures for
 $(\delta t,\gamma) = (1,1/2)$, $(\delta t,\gamma) \neq (\pm 1,\pm 1/2)$ and $(\delta t,\gamma) = (- 1,- 1/2)$.
 }
 \label{fig:SSH2D_PBC}
\end{figure*}

The rest of this paper is organized as follows.
In Sec.~\ref{sec:two-dimens-suschr}, we introduce the two-dimensional SSH model with nonreciprocal hoppings.
In addition, we show the energy spectra under the periodic boundary condition.
In Sec.~\ref{sec:mathbbz_4-berry-phas},
for the characterization of the real line gap topology,
we first introduce the $\mathbb{Z}_4$ Berry phase protected by generalized four-fold rotational symmetry.
We then show the explicit computation of the $\mathbb{Z}_4$ Berry phase in the present model.
In Sec.~\ref{sec:edge-spectr-cylind},
from the physical picture of the adiabatic connection,
we discuss the emergence of the edge states on the cylinder geometry.
Additionally, we characterize the point gap topology of edge spectra by the winding number.
In Sec.~\ref{sec:localized-states-at},
from the nontrivial point gap topology of the edge spectra,
we show the emergence of the corner states characterized by the pair of the $\mathbb{Z}_4$ Berry phase and the winding number.
In Sec.~\ref{sec:summary-}, we present a summery of this paper.

\section{Two-dimensional Su-Schrieffer-Heeger model with the nonreciprocal hoppings \label{sec:two-dimens-suschr}}
We consider the two-dimensional SSH model with the nonreciprocal hoppings [see Fig.~\ref{fig:SSH2D_PBC} (a)].
The red and blue lines denote the nonreciprocal hoppings.
Specifically, the hoppings along the red and blue arrows are $t_1 + \gamma_1$ and $t_2 + \gamma_2$, respectively.
The hoppings opposite to the red and blue arrows are $t_1 - \gamma_1$ and $t_2 - \gamma_2$, respectively.

Firstly, we consider the energy spectrum under the periodic boundary condition.
The Hamiltonian is
\begin{equation}
 H= \sum_{\vec{k}} \vec{c}^{\dagger}(\vec{k}) h(\vec{k}) \vec{c}(\vec{k}),
\end{equation}
with $\vec{c}(\vec{k}) = (c_1 (\vec{k}),c_2 (\vec{k}),c_3 (\vec{k}),c_4 (\vec{k}))^{T}$ being the vector of the annihilate operators.
Here, the explicit form of $h(\vec{k})$ is
\begin{equation}
 \label{eq:21}
  h(\vec{k})
  =
  \begin{pmatrix}
   q_{2\times2}(-k_y,k_y) &  q_{2\times2} (k_x,k_x) \\
   q_{2\times 2} (-k_x,-k_x) & q_{2\times 2} (k_y,-k_y)
  \end{pmatrix},
\end{equation}
with
\begin{equation}
 \label{eq:26}
  q_{2\times2} (k_1,k_2)
  =
  \begin{pmatrix}
   0 & t_1 - \gamma_1 + (t_2 - \gamma_2)e^{i k_1} \\
    t_1 + \gamma_1 + (t_2 + \gamma_2)e^{i k_2} &0
  \end{pmatrix}.
\end{equation}

This bulk Hamiltonian preserves the generalized four-fold rotational symmetry,
\begin{equation}
 \label{eq:2}
  U_4 h(\vec{k}) U_4^{-1} = h^{\dagger} (R_4 \vec{k}),
\end{equation}
with
\begin{equation}
 \label{eq:3}
  R_4 =
  \begin{pmatrix}
   0 & -1 \\
   1 & 0
  \end{pmatrix},
\end{equation}
and $U_4$ being a unitary operator
\begin{equation}
 \label{eq:4}
  U_4 =
  \begin{pmatrix}
   0 & 0 & 0 & 1 \\
   1 & 0 & 0 & 0 \\
   0 & 1 & 0 & 0 \\
   0 & 0 & 1 & 0
  \end{pmatrix}.
\end{equation}
The eigenvalues $E_{i}$ ($i=1$, $\cdots$, $4$) of $h(\vec{k})$ are
\begin{align}
  E_{1} (\vec{k})
  =-
  a(\vec{k})
  +
  \sqrt{b(\vec{k})},
 \quad
  & E_{2} (\vec{k})
  =
  a(\vec{k})
  +
  \sqrt{b(\vec{k})},   \nonumber \\
  E_{3} (\vec{k})
  =
  -a(\vec{k})
  -
  \sqrt{b(\vec{k})},
 \quad
  &E_{4} (\vec{k})
  =
  a(\vec{k})
  -
  \sqrt{b(\vec{k})},   \label{eq:30}
\end{align}
with
\begin{align}
 a(\vec{k})
 &=
 t_1^2 + t_2^2 - \gamma_1^2 - \gamma_2^2
 +(t_1 t_2 - \gamma_1 \gamma_2 ) (\cos k_x + \cos k_y), \label{eq:28} \\
 b(\vec{k})
 &=
 -(t_1 - \gamma_1)(t_1 + \gamma_1)(t_2 - \gamma_2) (t_2 + \gamma_2)(\cos k_x - \cos k_y)^2 \nonumber \\
 &+(t_1^2 + t_2^2 - \gamma_1^2 -\gamma_2^2 + (t_1 t_2 - \gamma_1 \gamma_2)(\cos k_x + \cos k_y))^2. \label{eq:29}
\end{align}

For the following discussions, we introduce $\delta t$ and $\delta \gamma$ as
\begin{align}
 t_1 = 1 + \delta t,  \quad
  &t_2 = 1 - \delta t, \label{eq:23} \\
 \gamma_1 = \frac{1}{2} + \delta \gamma, \quad
 &\gamma_2 = \frac{1}{2} - \delta \gamma.\label{eq:24}
\end{align}
In Figs.~\ref{fig:SSH2D_PBC} (b)-(d), the energy spectrum of the bulk Hamiltonian in Eq.~(\ref{eq:21}) is plotted
for $(\delta t , \delta \gamma) = (-0.7,0.1)$, $(0,0.1)$ and $(0.7,0.1)$.
There are real line gaps along the line $k_x = k_y$ for $-\delta t < \delta \gamma < \delta t$.

\section{$\mathbb{Z}_4$ Berry phase protected by generalized four-fold rotational symmetry \label{sec:mathbbz_4-berry-phas}}
We see the topological invariant to characterize the real line gap topology.
Specifically, we introduce the $\mathbb{Z}_4$ Berry phase
protected by generalized four-fold symmetry
in a two-dimensional momentum space.
The key idea originates from the quantized Berry phase
protected by the crystalline symmetry~\cite{Zak19890605,Hatsugai20110630,Kariyado20180615,Mizoguchi20190805,Kudo20190907,Araki20200109,Wakao20200330,Bunney20220110,Tsubota20220523}.
Indeed,
even in the discussion of the topology in  the two-dimensional systems,
the presence of rotational symmetry enables us to introduce the quantized Berry phase~\cite{Hatsugai20110630,Kariyado20180615,Kudo20190907,Mizoguchi20190805,Araki20200109,Wakao20200330,Bunney20220110}.
In addition, the notion of the complex Berry phase also plays an important role~\cite{Garrison19880328,Miniatura1990,Longhi20230213,Liang20130117,Tsubota20220523}.
Furthermore, in the presence of the crystalline symmetry,
the form of the topological invariant can be simplified to one determined from the data at high-symmetry points in the Brillouin zone.
This reduction is important from the view of the computational costs.

In this paper, we consider the topology of the bands at the left side of the real line gap.
The $\mathbb{Z}_4$ Berry phase is defined as follows.
Firstly, we define the $n_{\text{occ}} \times n_{\text{occ}} $ Berry connection as
\begin{equation}
 \label{eq:6}
 \vec{\bm{A}}(\vec{k})
  = -i
  \bm{\Phi}^{\dagger}(\vec{k})
  \frac{\partial}{\partial \vec{k}}
  \bm{\Psi}(\vec{k}),
\end{equation}
where
\begin{equation}
 \bm{\Psi} (\vec{k}) = (\vec{\psi}_1 (\vec{k}),\cdots,\vec{\psi}_{n_{\text{occ}}} (\vec{k})), \label{eq:7}
\end{equation}
and
\begin{equation}
  \bm{\Phi} (\vec{k}) = (\vec{\phi}_1 (\vec{k}),\cdots,\vec{\phi}_{n_{\text{occ}}} (\vec{k})), \label{eq:8}
\end{equation}
denote right and left eigenvectors in the momentum space, respectively.
Here, $n_{\text{occ}}$ is the number of the occupied states.
The $\mathbb{Z}_4$ Berry phase in the non-Hermitian system is
\begin{equation}
 \label{eq:5}
  \nu = \text{Tr} \hspace{1pt}
  \int_{L} \frac{d \vec{k}}{2\pi} \cdot \vec{\bm{A}}(\vec{k}),
\end{equation}
where $L$ denotes the paths between the high-symmetry points in the Brillouin zone [see Fig.\ref{fig:SSH2D_PBC} (e)].

To evaluate Eq.~(\ref{eq:5}), we consider the constraints on the eigenvectors and the Berry connection
arising from the presence of generalized four-fold symmetry.
In the presence of generalized four-fold symmetry, the constraints on right and left eigenvectors are
\begin{align}
 \bm{\Psi}^{\prime}(\vec{k}) = U_4\Phi (R_4^{-1} \vec{k}) \Theta (R_4^{-1}\vec{k}) \label{eq:9}\\
  \bm{\Phi}^{\prime}(\vec{k}) = U_4 \Psi (R_4^{-1} \vec{k}) \Theta (R_4^{-1}\vec{k}) \label{eq:10}
\end{align}
where
$ \Theta (\vec{k}) = \text{diag} (e^{i\theta_1 (\vec{k})} , \cdots , e^{i\theta_{n_{\text{occ}}} (\vec{k})})$
is the $n_{\text{occ}} \times n_{\text{occ}}$ matrix.
Here, the explicit forms of $\bm{\Psi}^{\prime} (\vec{k})$ and $\bm{\Phi}^{\prime} (\vec{k})$ are
\begin{align}
 \bm{\Psi}^{\prime} (\vec{k}) &= (\vec{\psi}_{1^{\ast}} (\vec{k}),\cdots,\vec{\psi}_{n_{\text{occ}}^{\ast}} (\vec{k})), \label{eq:15}\\
 \bm{\Phi}^{\prime} (\vec{k}) &= (\vec{\phi}_{1^{\ast}} (\vec{k}),\cdots,\vec{\phi}_{n_{\text{occ}}^{\ast}} (\vec{k})), \label{eq:16}
\end{align}
where $j^{\ast}$ ($j=1$, $\cdots$, $n_{\text{occ}}$) denotes the index with $E_{j^{\ast}} (\vec{k}) = (E_{j} ( \vec{k}))^{\ast} = E_{j} (R_4^{-1} \vec{k})$.
From the assumption of the presence of the real line gap,
the following holds:
if the band with the index $j$ is at the left side of the real line gap,
so is the band with the index $j^{\ast}$.
This indicates that there exist $\psi_j (\vec{k})$ ($\phi_j (\vec{k})$) and  $\psi_{j^{\ast}} (\vec{k})$ ($\phi_{j^{\ast}} (\vec{k})$)
as elements of the column vector $\bm{\Psi}^{\prime} (\vec{k})$ ($\bm{\Phi}^{\prime} (\vec{k})$).

From these constraints on the eigenvectors,
the Berry connection satisfies
\begin{align}
     \bm{A} (\vec{k})
  &= - i
  \Theta^{-1} (R_4^{-1} \vec{k})
  \left\{
   (\bm{\Psi}^{\prime} (R_4^{-1} \vec{k}))^{\dagger}
    \frac{\partial}{\partial \vec{k}}
   \bm{\Phi}^{\prime} (R_4^{-1} \vec{k})
  \right\}\Theta (R_4^{-1} \vec{k}) \nonumber
   \\
  &
   -i   \Theta^{-1} (R_4^{-1} \vec{k})
    \frac{\partial}{\partial \vec{k}}
   \Theta (R_4^{-1} \vec{k}).
   \label{eq:12}
\end{align}
By making use of Eq.~(\ref{eq:12}), we obtain
\begin{equation}
 \label{eq:13}
  \text{Re} \hspace{1pt}\nu =
  \frac{1}{2\pi}
  \sum_{i \leq n_{\text{occ}}}
  \left(
 \theta_i (M)
 -\theta_i (\Gamma)
    \right)
  = \frac{n }{4}
  \quad
 \text{mod}\hspace{3pt}1,
\end{equation}
where $n$ is $0$, $1$, $2$, or $3$.
Here, from Eq.~(\ref{eq:10}) and the biorthogonal normalization condition $\phi_{i}^{\ast} (\vec{k}^{\prime}) \psi_{i^{\prime}} (\vec{k}) = \delta_{ii^{\prime}} \delta (\vec{k} - \vec{k}^{\prime})$,
the explicit form of $\theta_i (\vec{k})$ is
\begin{equation}
 \label{eq:14}
  \theta_i (\vec{k})
  = -\text{arg}
  ((\vec{\psi}_{i^{\ast}} (R_4 \vec{k}))^{\dagger} U_4 \vec{\psi}_i (\vec{k})).
\end{equation}
The technical details regarding Eqs.~(\ref{eq:12}) and (\ref{eq:13}) are shown in Appendixes~\ref{sec:constr-eigenv-berry}, \ref{sec:quant-berry-phase}, and \ref{sec:simpl-mathbbz_4-berr}.
Even in the presence of gapless points on the integration path $L$
except the high-symmetry points $\Gamma$ and $M$ in Fig.~\ref{fig:SSH2D_PBC} (e),
by the continuous deformation of the integration path,
we can employ the $\mathbb{Z}_4$ Berry phase
for the characterization of the real line gap.
We show the detail and the example in Appendix~\ref{sec:eval-mathbbz_4-berry}.
This is the unique advantage of employing the $\mathbb{Z}_4$ Berry phase,
since the topological invariant such as the Chern number
can not be defined when there exists an exceptional point in the two-dimensional Brillouin zone.

We further show the explicit computation of the $\mathbb{Z}_4$ Berry phase for the present model.
We note that we focus on the real line gap at the red line in Figs.~\ref{fig:SSH2D_PBC} (b) and (d).
At the high symmetry points $\Gamma$ and $M$,
the explicit forms of the right eigenvectors of $E_{1}$
which are the isolated energy spectrum at the left side in Fig.~\ref{fig:SSH2D_PBC} (a)
are
\begin{align}
 & \vec{\psi}_{1} (\Gamma)
  =\left(
    -\frac{1}{2\sqrt{3}},
    \frac{1}{2},
    -\frac{1}{2\sqrt{3}},
    \frac{1}{2}
   \right)^T,   \label{eq:1}\\
 & \vec{\psi}_{1} (M)
  =\left(
 -\frac{\sqrt{\delta t^2 - \delta \gamma^2}}{2(\delta t- \delta\gamma)},
 \frac{1}{2},
 -\frac{\sqrt{\delta t^2 - \delta \gamma^2}}{2(\delta t- \delta\gamma)},
 \frac{1}{2}
 \right)^T, \label{eq:11}
\end{align}
respectively.
Thus, the explicit forms of $\theta_{1} (\vec{k})$ at $\Gamma$ and $M$ are
\begin{align}
   \theta_{1}(\Gamma)
  =-\text{arg} \left(-\frac{1}{\sqrt{3}}\right),  \quad
  \theta_{1} (M)
  =-\text{arg}\left(- \frac{\sqrt{\delta t^2-\delta \gamma^2}}{\delta t - \delta \gamma}\right). \label{eq:22}
\end{align}
In Fig.\ref{fig:SSH2D_PBC} (f), we show the color map of the $\mathbb{Z}_4$ phase of the band $E_{1} (\vec{k})$ in Eq.~(\ref{eq:13}).
The $\mathbb{Z}_4$ Berry phase takes values of $\text{Re}\hspace{1pt}\nu = 1/2$ (yellow area) and $\text{Re}\hspace{1pt}\nu = 0$ (blue area).
The white area shows the existence of the exceptional rings around the high-symmetry point $M$.
We note that, in the present model,
there are only two topological phases ($\text{Re}\hspace{1pt}\nu = 0$, $1/2$)
whereas the Berry phase is quantized in $\mathbb{Z}_4$.
In Appendix~\ref{sec:mathbbz_4-topol-phas},
we show an example of the emergence of four topological phases in the different toy model.

Before closing this discussion,
we address the adiabatic connection of the $\mathbb{Z}_4$ Berry phase.
Since the $\mathbb{Z}_4$ Berry phase does not change as long as the real line gap is closed,
the system can be adiabatically deformed into a certain limit that has the same Berry phase.
In the present model, the phase with $\text{Re}\hspace{1pt}\nu = 0$ is connected to the limit of $\delta t = 1$ and $\delta \gamma = 1/2$ [see Fig.~\ref{fig:SSH2D_PBC} (g)],
while the phase with $\text{Re}\hspace{1pt}\nu  = 1/2$ is connected to the limit of $\delta t = -1$ and $\delta \gamma = -1/2$ [see Fig.~\ref{fig:SSH2D_PBC} (g)].
This insight is essential to understand the relation between edge states and the $\mathbb{Z}_4$ Berry phase.

\begin{figure*}[t]
 \centering
 \includegraphics[trim={0cm 0cm 0cm 0cm},width =0.9\hsize]{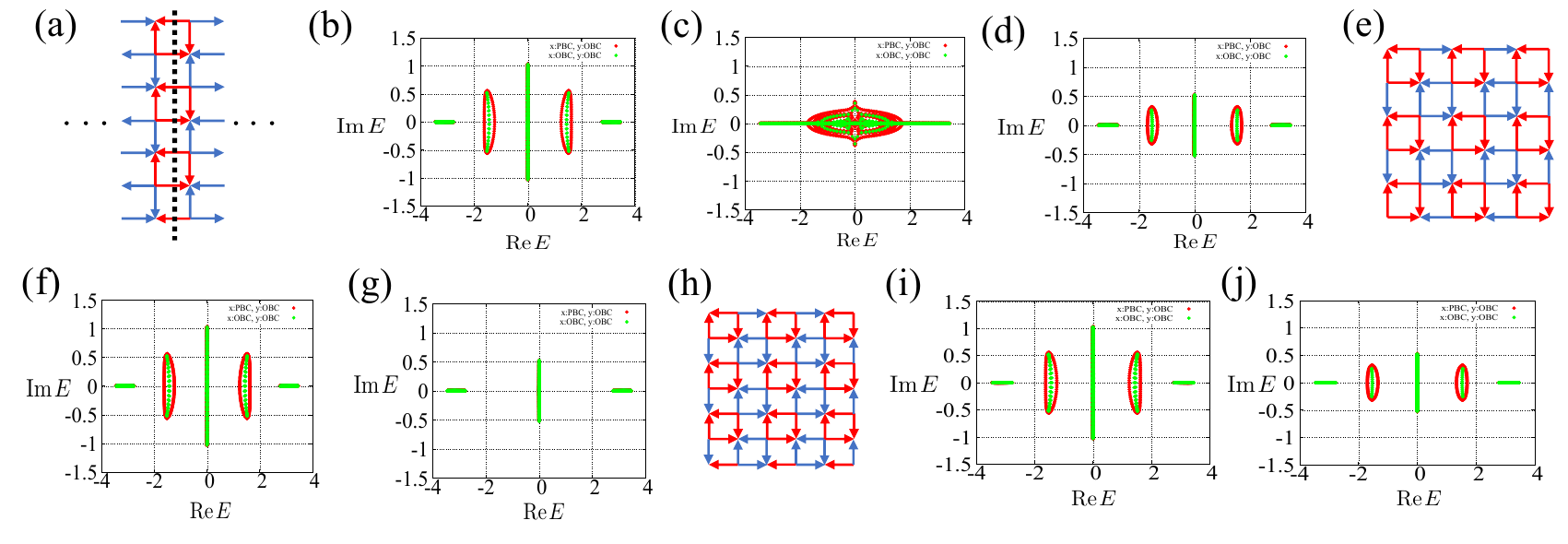}
 \caption{(Color online)
 (a) Schematic figure for the boundary shape of the two-dimensional SSH model with nonreciprocal hoppings when considering the cylinder geometry.
 By breaking inversion symmetry, we avoid vanishing of the winding number.
 The dashed line shows the center of the generalized mirror plane.
 (b)-(d) The red (green) dots show the energy spectrum with the geometry in (a) under the periodic (open) boundary condition along the $x$ direction for several parameters sets of $(\delta t , \delta \gamma)$.
 (b), (c), and (d) show the data for $(\delta t , \delta \gamma) = (-0.7,0.1)$, $(0,0.1)$, and $(0.7,0.1)$, respectively.
 (e) Schematic illustration of the non-Hermitian two-dimensional SSH model with even number of sites along both $x$ and $y$ directions.
 (f)-(g) The red (green) dots show energy spectra under the periodic (open) boundary condition along the $x$ direction with the geometry in (e).
 (f) and (g) show the energy spectra on the complex plane for $(\delta t , \delta \gamma) = (-0.7,0.1)$ and $(0.7,0.1)$, respectively.
 (h) Schematic illustration of the non-Hermitian two-dimensional SSH model with even (odd) number of sites along the $x$ ($y$) direction.
 (i)-(j) The red (green) dots represent energy spectra under the periodic (open) boundary condition along the $x$ direction with the geometry in (h).
 (i) and (j) show the energy spectra on the complex plane for $(\delta t , \delta \gamma) = (-0.7,0.1)$, $(0.7,0.1)$, respectively.
 }
 \label{fig:SSH2D_OBC}
\end{figure*}

\section{Edge spectrum on the cylinder geometry \label{sec:edge-spectr-cylind}}
Before going to the numerical result of the edge spectra,
to get the insight of the relation between the $\mathbb{Z}_4$ Berry phase and edge states,
we consider edge states originating from the dimers at the two limits, i.e., $(\delta t , \delta \gamma) = (1,1/2)$ and $(\delta t, \delta \gamma) = (-1,-1/2)$.
Indeed,
as shown in the previous work~\cite{Mizoguchi20190805,Araki20200109},
considering the adiabatic connection is
a key to explain the correspondence
between the Berry phase and the boundary modes,
since it gives the simple real-space picture of the state.
By the analogous approach, we consider the two certain limits in the present model [see Fig.~\ref{fig:SSH2D_PBC} (g)].
Specifically, for $(\delta t , \delta \gamma) = (1,1/2)$, there exist dimers originating from the cut of the unit cell on the right and bottom sides,
while there exist dimers on the left and top sides for $(\delta t, \delta \gamma) = (-1,-1/2)$.
These dimers result in the emergence of edge states.
By combining with the discussion of the adiabatic connection of the $\mathbb{Z}_4$ Berry phase,
$\text{Re}\hspace{1pt}\nu = 1/2$ ($0$) predicts the emergence of the edge states at left and top (right and bottom) sides.
Thus, the $\mathbb{Z}_4$ Berry phase indicates the position of edge states~\cite{footnote1}.

Once the profile of edge states are understood by the $\mathbb{Z}_4$ Berry phase,
we can then argue the characterization of the skin effect of the edge spectra.
To be more concrete,
the non-zero winding number of the edge spectrum provides the emergence of the hybrid higher-order skin topological modes.
The explicit form of the winding number of the edge spectra is
\begin{equation}
 \label{eq:17}
  w = \int^{\pi}_{-\pi} \frac{d k}{2\pi i}\hspace{1pt} \frac{\partial}{\partial k} \ln \det (h_{1\text{D}}(k) - E_{\text{ref}} 1_{N \times N}),
\end{equation}
where $h_{1\text{D}}(k)$ and $E_{\text{ref}}$ denote the Hamiltonian on the cylinder geometry and the reference energy, respectively.
However, in the presence of the conventional inversion symmetry [see Eq.~(\ref{eq:2})],
the total of the winding number vanishes~\cite{Okugawa20210521}.
Specifically,
the one-dimensional Hamiltonian preserves conventional inversion symmetry~\cite{Okugawa20210521},
\begin{equation}
\label{eq:52}
  U_I h_{1\text{D}}(k) U_I^{-1} = h_{1\text{D}}(-k),
\end{equation}
with $U_I$ being the unitary matrix ($U_I^2 = 1_{N \times N}$).
The presence of this symmetry enables us to evaluate the winding number,
\begin{equation}
\label{eq:57}
  \begin{split}
   w &= \int^{\pi}_{-\pi} \frac{dk}{2\pi i} \frac{\partial}{\partial k}\ln \det (h_{1\text{D}}(k)-E_{\text{ref}} 1_{N\times N})   \\
    &= \int^{\pi}_{-\pi} \frac{dk}{2\pi i} \frac{\partial}{\partial k}\ln \det (U_I (h_{1\text{D}}(-k)-E_{\text{ref}}1_{N\times N})U_{I}^{-1})   \\
   &= -\int^{\pi}_{-\pi} \frac{dk}{2\pi i} \frac{\partial}{\partial k}\ln \det (h_{1\text{D}}(k)-E_{\text{ref}1_{N\times N}})   \\
   &= -w.
  \end{split}
\end{equation}
Here, we have used Eq.~(\ref{eq:52}) and the relation $\det (AB) = \det (BA)$.
Thus, in the presence of the inversion symmetry, the total of the winding number vanishes.
To avoid this problem, we consider the system in the absence of inversion symmetry
by setting the appropriate edges [see Fig.~\ref{fig:SSH2D_OBC} (a)].
The trivial case of edge spectra is discussed in Sec.~\ref{sec:localized-states-at}.

 \begin{figure*}[t]
 \centering
  \centering
\includegraphics[trim={0cm 0cm 0cm 0cm},width =0.8\hsize]{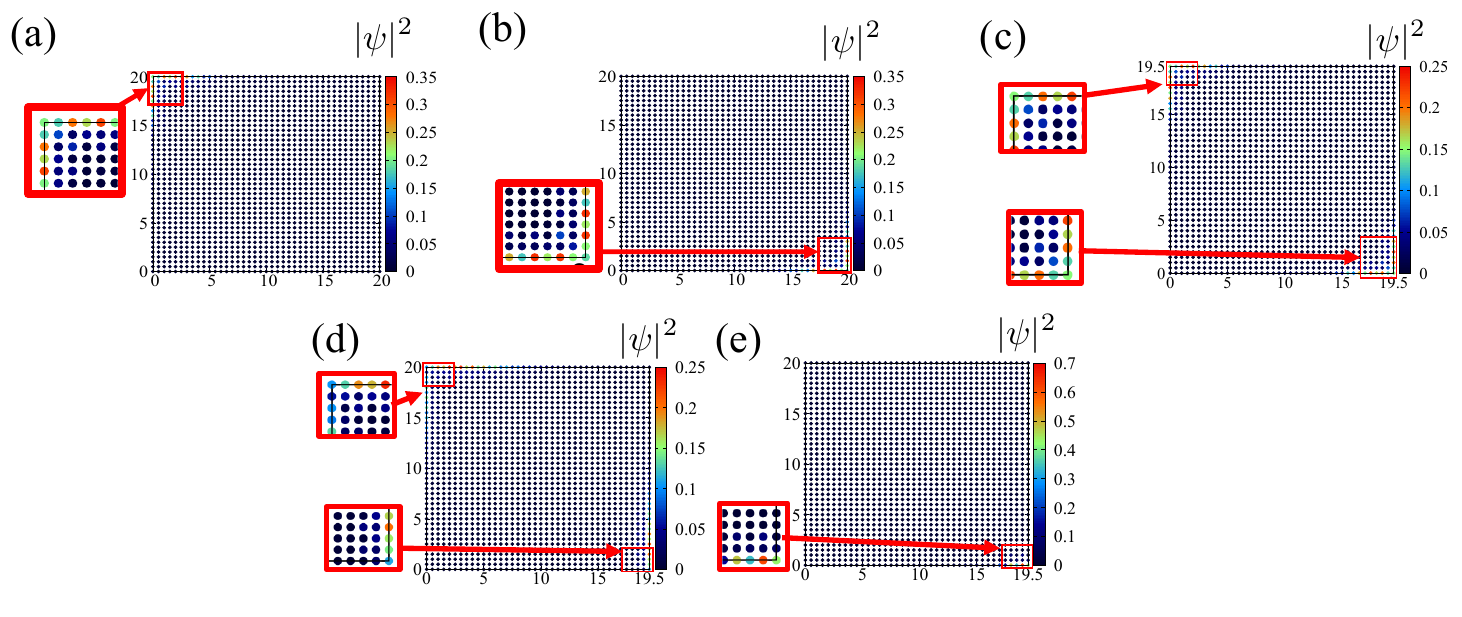}
 \caption{(Color online)
  (a) and (b) show the distribution of the right eigenvector in the real space of corner states with the geometry in Fig.~\ref{fig:SSH2D_OBC} (a) for $(\delta t , \delta \gamma, E) = (-0.7,0.1,-1.51302+0.480218i)$ and $(0.7,0.1,-1.54824+0.260155i)$, respectively.
  (c) shows the distribution of the eigenstate with the geometry in Fig.~\ref{fig:SSH2D_OBC} (e) for $(\delta t, \delta \gamma, E) = (-0.7,0.1,-1.50723526-0.47764771i)$.
  (d) and (e) show the distributions of the eigenstates with the geometry in Fig.~\ref{fig:SSH2D_OBC} (h) for $(\delta t, \delta \gamma, E) = (-0.7,0.1,-1.52665387-0.52222989i)$ and $(0.7,0.1,-1.5395-0.193811i)$, respectively.
  }
 \label{fig:SSH2D_vec}
 \end{figure*}

We note that, in the present model,
we can also define the $\mathbb{Z}_2$ indicator
with generalized mirror symmetry
that predicts the emergence of the skin effect~\cite{Okugawa20210521} [see Fig.~\ref{fig:SSH2D_OBC} (a)].
In the presence model, the Hamiltonian preserves the generalized mirror symmetry, namely, it satisfies
\begin{equation}
\label{eq:53}
  U_M h_{1\text{D}}(k) U_M^{-1} = h^{\dagger}_{1\text{D}} (-k),
\end{equation}
 where $U_M$ denotes the unitary operator.
 The presence of this symmetry enables us to define the $\mathbb{Z}_2$ indicator~\cite{Okugawa20210521},
 \begin{equation}
  \label{eq:54}
   \mu := (-1)^w = \prod_{k=0,\pi}\text{sgn} (\det (h_{1\text{D}}(k)- E_{\text{ref}}1_{N \times N})).
 \end{equation}
Here, we have assumed that the reference energy $E_{\text{ref}}$ is real.
The nontrivial value of the indicator (i.e., $\mu = -1$) predicts the emergence of the skin effect, since it guarantees that $w$ is nonzeo.
In the present system, the pair of the $\mathbb{Z}_4$ Berry phase and this $\mathbb{Z}_2$ indicator predicts the emergence of the hybrid higher-order skin topological modes.
However, $\mu$ cannot predict the position of the hybrid higher-order skin topological modes since we cannot see the difference between $w=1$ and $w=-1$.

Figures~\ref{fig:SSH2D_OBC} (b), (c), and (d) display
energy spectra under the periodic (open) boundary condition along $x$ ($y$) direction
for $(\delta t, \delta \gamma) = (-0.7,0.1)$, $(0,0.1)$, and $(0.7,0.1)$, respectively.
Here, we set the number of the unit cells as $20$.
Thus, the total number of the lattice sites are $82$.
There exists the energy spectrum of edge states
where eigenvectors are localized at top and left [bottom and right] side for
$(\delta t, \delta \gamma) = (-0.7,0.1)$ $[(0.7,0.1)]$.
Here, we note that there may exist the energy spectrum of edge states
even in the presence of gapless states.
Thus, the emergence of all edge states is not predicted by the $\mathbb{Z}_4$ Berry phase.

Before closing this section, we see energy spectra in the presence of conventional inversion symmetry.
Specifically, we consider the system with conventional inversion symmetry by setting appropriate boundaries [see Fig.~\ref{fig:SSH2D_OBC} (e)].
Figure~\ref{fig:SSH2D_OBC} (f) displays energy spectra under the periodic (open) boundary condition along the $x$ ($y$) direction for $(\delta t, \delta \gamma) = (-0.7,0.1)$.
At first sight, this numerical result of edge spectra looks similar to the one in Fig.~\ref{fig:SSH2D_OBC} (b).
However,
while there are two edge spectra in the absence of conventional inversion symmetry [see Fig.~\ref{fig:SSH2D_OBC} (b)],
there are four edge spectra in the presence of conventional inversion symmetry [see Fig.~\ref{fig:SSH2D_OBC} (f)],
each of which is doubly degenerated.
In the presence of the inversion symmetry,
one of edge spectra at the left side in Fig.~\ref{fig:SSH2D_OBC} (f) is with $w=1$.
Another one at the left side in Fig.~\ref{fig:SSH2D_OBC} (f) is with $w=-1$.
Then, the total of the winding number with conventional inversion symmetry vanishes.
Therefore, in the conventional inversion symmetry, we cannot distinguish this case from the trivial loop of edge spectra.
However, the corner states emerge even in the case for $w=0$.
The detail is discussed in Sec.~\ref{sec:localized-states-at}.

\section{Localized states at the corner \label{sec:localized-states-at}}
Finally, we demonstrate that the system hosts corner skin modes.
We consider the $1681$-site system with the open boundary condition in both directions [see Fig.~\ref{fig:SSH2D_PBC} (g)].
In Figs.~\ref{fig:SSH2D_OBC} (b) and (d),
we see the emergence of the corner skin modes arising from the edge states at the left top (right bottom) corner for $\delta t = -0.7$ and $0.7$ [see Figs.~\ref{fig:SSH2D_vec} (a) and (b)].
In addition, we note that, in the present model,
the conventional corner states in the Hermitian limit cannot be predicted by the $\mathbb{Z}_4$ Berry phase
since such states may be buried in the bulk at zero modes.

Combining the above numerical result and the topological invariants,
we conclude that the following bulk-boundary correspondence between the pair of topological number $(\text{Re}\hspace{1pt}\nu,w)$ and the hybrid higher-order skin topological modes holds.
Specifically,
$(\text{Re}\hspace{1pt}\nu,w) = (1/2,-1)$ indicates the emergence of the hybrid higher-order skin topological modes at the left top corner,
while $(\text{Re}\hspace{1pt}\nu,w) = (0,1)$ indicates the emergence of the hybrid higher-order skin topological modes at the right bottom corner.
We note that the case for $(\text{Re}\hspace{1pt}\nu,w) = (1/2,1)$ and $(0,-1)$ is discussed in Appendix~\ref{sec:corn-stat-char}.
This correspondence
between the topological invariants and hybrid higher-order skin topological modes
originates from the physical picture of the adiabatic connection.

\begin{figure}[h]
 \centering
  \centering
\includegraphics[trim={0cm 0cm 0cm 0cm},width =\hsize]{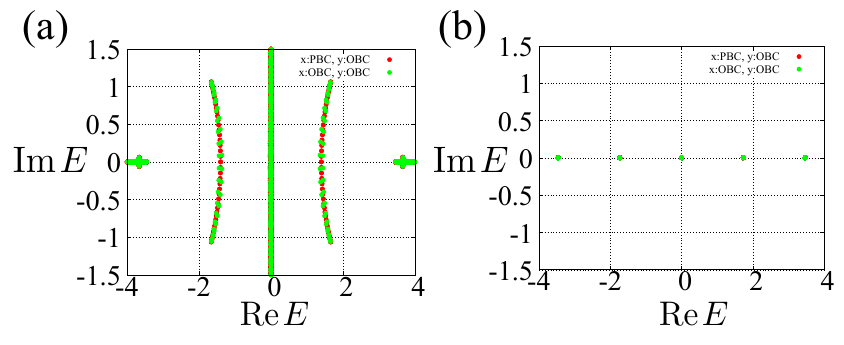}
 \caption{(Color online)
 (a) and (b) show the energy spectra for $(\delta t , \delta \gamma) = (-1,1/2)$, and $(-1,-1/2)$, respectively.
 The red (green) dots show the energy spectrum under the periodic (open) boundary condition along the $x$ direction.
 }
 \label{fig:SSH2D_Trivial}
\end{figure}

We note that the present model also hosts the trivial point gap topology.
To show examples in the present model,
we plot
the energy spectra under the periodic (open) boundary condition along $x$ direction
for $(\delta t, \delta \gamma) = (-1,1/2)$ and $(-1,-1/2)$ [see Fig.~\ref{fig:SSH2D_Trivial}].
In Fig.~\ref{fig:SSH2D_Trivial}, there are trivial loops of the edge spectra between the bulk edge spectra.
For these parameters, there does not exists the emergence of hybrid higher-order skin topological modes.

 Before summarizing these results, we show the emergence of corner states with odd number of sites along both $x$ and $y$ directions.
 Here, we show that the corner modes emerge in the different shape of the boundary as well.

 First, we see the system with even number of sites along both $x$ and $y$ directions,
 where the total sites are 1600-sites [see Fig.~\ref{fig:SSH2D_OBC} (e)].
 This system respects conventional inversion symmetry.
 While there exist in-gap states arising from the nontrivial point gap topology of edge spectra for $(\delta t , \delta \gamma) =(-0.7,0.1)$,
 there does not exist such states for $(\delta t , \delta \gamma) =(0.7,0.1)$ due to the absence of edge spectra on the cylinder geometry [see Figs.~\ref{fig:SSH2D_OBC} (f) and (g)].
 In Fig.~\ref{fig:SSH2D_vec} (c), we observe the emergence of eigenstates localized at both left top and right bottom corners arising from the localized states at all edges.
 Here, as discussed above, we note that the topology of these point gaps cannot be characterized by the winding number on the cylinder geometry due to the presence of inversion symmetry.

 Second, we see the system with even (odd) number of sites along the $x$ ($y$) direction,
 where the total sites are 1640-sites[see Fig.~\ref{fig:SSH2D_OBC} (h)].
 Figures~\ref{fig:SSH2D_OBC} (i) and (j) display the energy spectra on the complex plane for $(\delta t , \delta \gamma) = (-0.7,0.1)$, $(0.7,0.1)$, respectively.
 In these figures, there exist the energy spectra of corner states.
 While the corner states for $(\delta t , \delta \gamma) = (-0.7,0.1)$ are localized at left top and right bottom corners,
 such states for $(\delta t , \delta \gamma) = (0.7,0.1)$ are localized at the right bottom corner.
 The former state is provided by the edge states localized at left, top, and right boundaries.
 The latter state is provided by the edge states localized at the bottom boundary.
 Here, we note that these corner states can be also characterized by the pair of the $\mathbb{Z}_4$ Berry phase and the winding number under the periodic (open) boundary condition along the $x$ ($y$) direction.

\section{Summary \label{sec:summary-}}
We have realized the hybrid higher-order skin topological modes in two-dimensional SSH model with nonreciprocal hoppings.
Specifically, we characterize the real line gap topology by introducing the $\mathbb{Z}_4$ Berry phase protected by generalized four-fold rotational symmetry.
Additionally, from the discussion of the adiabatic connection, we show the relation between the position of edge states and the topological invariant.
Moreover, from the nontrivial point gap arising by the edge spectra, we demonstrate that the present model hosts the localized states at the corner.
In particular, the position of these localized states are predicted by the pair of the $\mathbb{Z}_4$ Berry phase and the winding number.

\section{Acknowledgment}
The author thanks T. Mizoguchi for fruitful discussions.
This work is supported by JST, the
establishment of university fellowships towards the creation of science technology innovation,
Grant No.~JPMJFS2106.

\appendix

\section{Constraints on the eigenvectors and the Berry connection by generalized four-fold rotational symmetry\label{sec:constr-eigenv-berry}}
Here, we show the details of the constraint on the eigenvectors and the Berry connection arising from the presence of generalized four-fold rotational symmetry.
First, we consider the constraints from generalized four-fold rotational symmetry~\cite{Tsubota20220523,Chen20220224}.
From the constraint on the Hamiltonian, the eigenvalue equation is rewritten as
\begin{equation}
\label{eq:31}
  \begin{split}
   H(\vec{k}) \psi_{n}(\vec{k}) &= E_n(\vec{k}) \psi_{n}(\vec{k}) \\
    H^{\dagger}(R_4 \vec{k}) U_4 \psi_{n}(\vec{k}) &= E_n(\vec{k}) U_4 \psi_{n}(\vec{k}),
  \end{split}
\end{equation}
with $n$ being the band index.
Thus, the constraint on the right eigenvectors are
\begin{equation}
\label{eq:32}
  \phi_{j^{\ast}} (R_4 \vec{k}) = e^{i \theta_j (\vec{k})} U_4 \psi_j (\vec{k}).
\end{equation}
By the same approach, we also obtain the constraint on the left eigenvectors
\begin{equation}
\label{eq:33}
  \psi_{j^{\ast}} (R_4 \vec{k}) = e^{i \theta_j (\vec{k})} U_4 \phi_j (\vec{k}).
\end{equation}
Here, the phase factor in Eq.~(\ref{eq:33}) is determined from the biorthogonal normalization condition $\phi_{j}^{\ast} (\vec{k}^{\prime}) \psi_{j^{\prime}} (\vec{k}) = \delta_{jj^{\prime}} \delta (\vec{k} - \vec{k}^{\prime})$ [see Eq.~(\ref{eq:14})].
Thus, we obtain the relation in Eqs.~(\ref{eq:32}) and (\ref{eq:33}).

Second, we show the relation in Eq.~(\ref{eq:12}).
From the constraints on the eigenvectors, the Berry connection is rewritten as
\begin{equation}
\label{eq:34}
  \begin{split}
   &    \vec{\bm{A}}(\vec{k})  \\
   &    = -i \left\{ \Theta^{-1} (R_4^{-1} \vec{k}) (\bm{\Psi}^{\prime})^{\dagger} (R_4^{-1} \vec{k}) U_4^{-1} \right\} \frac{\partial}{\partial \vec{k}}  \left\{ U_4 \bm{\Phi}^{\prime} (R_4^{-1} \vec{k}) \Theta (R_4^{-1} \vec{k}) \right\} \\
   &    = -i  \Theta^{-1} (R_4^{-1} \vec{k})    \left( (\bm{\Psi}^{\prime})^{\dagger} (R_4^{-1} \vec{k})\frac{\partial}{\partial \vec{k}}  \bm{\Phi}^{\prime} (R_4^{-1} \vec{k}) \right) \Theta (R_4^{-1} \vec{k}) \\
   &    \quad -i  \Theta^{-1} (R_4^{-1} \vec{k})   \frac{\partial}{\partial \vec{k}}  \Theta (R_4^{-1} \vec{k}).
  \end{split}
\end{equation}
Thus, we obtain Eq.~(\ref{eq:12}).

\section{Quantization of the Berry phase with generalized four-fold rotational symmetry\label{sec:quant-berry-phase}}
\begin{figure}[t]
 \centering
  \centering
\includegraphics[trim={0cm 0cm 0cm 0cm},width =0.4\hsize]{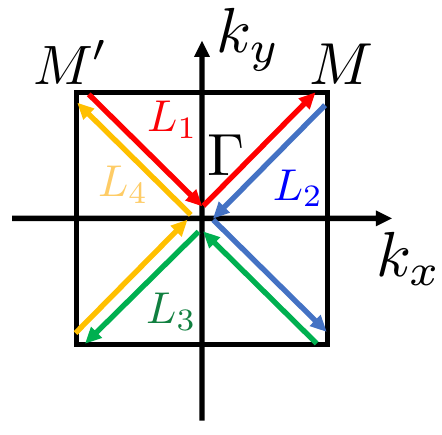}
 \caption{(Color online)
 Sketch of the Brillouin zone and the path of the integration in Eq.~(\ref{eq:7}).
 }
 \label{fig:SSH2D_Berry}
\end{figure}

We show the details of the quantization of the Berry phase.
In order to show the $\mathbb{Z}_4$ quantization of $\text{Re}\hspace{1pt} \nu$,
we consider the sum of the Berry phase defined as a contour integral along four paths ($L_1$-$L_4$) in Fig.~\ref{fig:SSH2D_Berry},
\begin{equation}
\label{eq:35}
  \nu_{\text{all}} = \sum_{i = 1}^4 \nu_i,
\end{equation}
with
\begin{equation}
\label{eq:36}
  \nu_{i} = \text{Tr} \int_{L_i} \frac{d\vec{k}}{2\pi} \cdot \vec{\bm{A}}(\vec{k}).
\end{equation}
Here, the path $L_i$ connects high-symmetry points in Fig.~\ref{fig:SSH2D_Berry}.
We note that $L_1$ is identical to $L$ in Eq.~(\ref{eq:5}).
Since $\nu_{\text{all}}$ is obtained from the integration along the closed path, we obtain $\nu_{\text{all}} = n^{\prime}$ with an integer $n^{\prime}$.
From the constraints on the Berry connection, $\nu_{2}$ can be rewritten as
\begin{equation}
\label{eq:37}
  \begin{split}
       \nu_{2}
   &    =\text{Tr} \int_{L_2} \frac{d \vec{k}}{2\pi} \cdot
   \left\{
   -i  \Theta^{-1} (R_4^{-1} \vec{k})    \left( (\bm{\Psi}^{\prime})^{\dagger} (R_4^{-1} \vec{k}) \right. \right. \\
   &    \left. \left.
   \quad \times
   \frac{\partial}{\partial \vec{k}}
    \bm{\Phi}^{\prime} (R_4^{-1} \vec{k}) \right) \Theta (R_4^{-1} \vec{k})
   \right\} \\
   &    \quad +\text{Tr} \int_{L_2} \frac{d \vec{k}}{2\pi} \cdot
   \left\{
    -i  \Theta^{-1} (R_4^{-1} \vec{k})   \frac{\partial}{\partial \vec{k}}  \Theta (R_4^{-1} \vec{k})
   \right\} \\
   &    =\text{Tr} \int_{L_2} \frac{d \vec{k}}{2\pi} \cdot
     \left\{
   -i     \left( \bm{\Psi}^{\dagger} (R_4^{-1} \vec{k})\frac{\partial}{\partial \vec{k}}  \bm{\Phi} (R_4^{-1} \vec{k}) \right)
   \right\} \\
   &     \quad + \sum_{i=1}^{n_{\text{occ}}} \int_{L_2} \frac{d \vec{k}}{2\pi i} \cdot \frac{\partial}{\partial \vec{k}} \ln e^{i\theta_i (R_4^{-1}\vec{k})}  \\
   &    =\text{Tr} \int_{L_1} \frac{d \vec{k}}{2\pi} \cdot
     \left\{
   -i     \left( \bm{\Psi}^{\dagger} (\vec{k})\frac{\partial}{\partial \vec{k}}  \bm{\Phi} ( \vec{k}) \right)
   \right\}
   + n_2,
  \end{split}
\end{equation}
where $n_2$ is an integer.
Here, we have used the relation $\text{Tr} (ABC) = \text{Tr} (CAB)$.
Additionally,
we have used
\begin{equation}
\label{eq:38}
  \text{Tr} \left((\bm{\Psi}^{\prime})^{\dagger} (R_4^{-1} \vec{k})\frac{\partial}{\partial \vec{k}}  \bm{\Phi}^{\prime} (R_4^{-1} \vec{k})    \right)
  =\text{Tr} \left(\bm{\Psi}^{\dagger} (R_4^{-1} \vec{k})\frac{\partial}{\partial \vec{k}}  \bm{\Phi} (R_4^{-1} \vec{k})    \right),
\end{equation}
since
$\bm{\Phi}^{\prime} (\vec{k})$ ($\bm{\Psi}^{\prime} (\vec{k})$) is obtained by replacing the order of the column vectors in $\bm{\Phi} (\vec{k})$ ($\bm{\Psi}(\vec{k})$).
This relation is originated from the assumption that the bands $E_j$ and $E_{j^{\ast}}$ are located
at the left side of the real line gap in the complex plane of the energy distribution.

By the same approach, we obtain the equations,
\begin{align}
  \nu_3 & =  \text{Tr} \int_{L_1} \frac{d\vec{k}}{2 \pi } \cdot \vec{\bm{A}}(\vec{k}) +  n_3 ,  \label{eq:56} \\
  \nu_4 & =  \text{Tr} \int_{L_1} \frac{d\vec{k}}{2 \pi } \cdot      \left\{
   -i     \left( \bm{\Psi}^{\dagger} (\vec{k}) \frac{\partial}{\partial \vec{k}}  \bm{\Phi} ( \vec{k}) \right)
   \right\}
 +  n_4,  \label{eq:55}
\end{align}
with integers $n_3$ and $n_4$.
By substituting Eqs.~(\ref{eq:38}), (\ref{eq:56}), and (\ref{eq:55}) into Eq.~(\ref{eq:35}), we obtain
\begin{equation}
\label{eq:39}
  \begin{split}
   \nu_{\text{all}} &= 2\text{Tr} \int_{L_1} \frac{d\vec{k}}{2 \pi } \cdot \vec{\bm{A}}(\vec{k}) \\
   & \quad +2\text{Tr} \int_{L_1} \frac{d\vec{k}}{2 \pi } \cdot
   \left\{
   -i     \left( \bm{\Psi}^{\dagger} (\vec{k})\frac{\partial}{\partial \vec{k}}  \bm{\Phi} ( \vec{k}) \right)
   \right\} + \sum_{i=2}^4 n_i \\
   &= 2\text{Tr} \int_{L_1} \frac{d\vec{k}}{2 \pi } \cdot \vec{\bm{A}}(\vec{k}) \\
   & \quad +2\text{Tr} \int_{L_1} \frac{d\vec{k}}{2 \pi } \cdot      \left\{
   -i   \left(  \frac{\partial}{\partial \vec{k}}  \bm{\Phi} ( \vec{k}) \right)^T \bm{\Psi}^{\ast}(\vec{k})
   \right\} + \sum_{i=2}^4 n_i \\
   &= 2\text{Tr} \int_{L_1} \frac{d\vec{k}}{2 \pi } \cdot \vec{\bm{A}}(\vec{k})
   +2\text{Tr} \int_{L_1} \frac{d\vec{k}}{2 \pi } \cdot \vec{\bm{A}}^{\ast}(\vec{k})
    + \sum_{i=2}^4 n_i \\
   &= 4\text{Re} \left(\text{Tr} \int_{L_1} \frac{d\vec{k}}{2 \pi } \cdot \vec{\bm{A}}(\vec{k})\right)
   + \sum_{i=2}^4 n_i.
  \end{split}
\end{equation}
Here, we have used the integration by parts and the relation $\text{Tr}A = \text{Tr} A^{T}$.
Therefore, form Eq.~(\ref{eq:39}) and $\nu_{\text{all}} =  n$ ($n \in \mathbb{Z}$), we obtain the quantization of the Berry phase in Eq.~(\ref{eq:13}).

\section{Simplification of the $\mathbb{Z}_4$ Berry phase\label{sec:simpl-mathbbz_4-berr}}
We show that the $\mathbb{Z}_4$ Berry phase can be computed from the data at the high-symmetry points $\Gamma$ and $M$
from the evaluation in Eq.~(\ref{eq:55}),
\begin{equation}
\label{eq:40}
  \begin{split}
   \nu &=
   \text{Tr}
   \int_{M^{\prime} \rightarrow \Gamma} \frac{d \vec{k}}{2 \pi } \cdot \vec{\bm{A}} (\vec{k})
   +\text{Tr}\int_{\Gamma \rightarrow M} \frac{d \vec{k}}{2 \pi } \cdot \vec{\bm{A}} (\vec{k})\\
   &=
   \text{Tr}   \int_{M^{\prime} \rightarrow \Gamma} \frac{d \vec{k}}{2 \pi } \cdot \vec{\bm{A}} (\vec{k}) \\
   & \quad +\text{Tr}
   \int_{\Gamma \rightarrow M}
   \frac{d \vec{k}}{2 \pi } \cdot
     \left\{
   -i     \left( \bm{\Psi}^{\dagger} (R_4^{-1} \vec{k})\frac{\partial}{\partial \vec{k}}  \bm{\Phi} (R_4^{-1} \vec{k}) \right)
   \right\} \\
    & \quad + \sum_{i=1}^{n_{\text{occ}}} \int_{\Gamma \rightarrow M} \frac{d \vec{k}}{2\pi i} \cdot \frac{\partial}{\partial \vec{k}} \ln  e^{i\theta_i (R_4^{-1} \vec{k})}  \\
   &=
   \text{Tr}   \int_{M^{\prime} \rightarrow \Gamma} \frac{d \vec{k}}{2 \pi } \cdot \vec{\bm{A}} (\vec{k})
    +\text{Tr}
   \int_{\Gamma \rightarrow M^{\prime}}
   \frac{d \vec{k}}{2 \pi } \cdot
     \left\{
   -i     \left( \bm{\Psi}^{\dagger} ( \vec{k})\frac{\partial}{\partial \vec{k}}  \bm{\Phi} ( \vec{k}) \right)
   \right\} \\
   & \quad  +\sum_{i=1}^{n_{\text{occ}}} \int_{\Gamma \rightarrow M^{\prime}} \frac{d \vec{k}}{2\pi i} \cdot \frac{\partial}{\partial \vec{k}} \ln  e^{i \theta_i (\vec{k})}   \\
   &=
   \text{Tr}   \int_{M^{\prime} \rightarrow \Gamma} \frac{d \vec{k}}{2 \pi } \cdot \vec{\bm{A}} (\vec{k})
   + i n_{\text{occ}}
    -\text{Tr}
   \int_{M^{\prime} \rightarrow \Gamma}
   \frac{d \vec{k}}{2 \pi } \cdot
   \vec{\bm{A}}^{\ast} (\vec{k}) \\
   & \quad +\frac{1}{2\pi}\sum_{i=1}^{n_{\text{occ}}} (\theta (M^{\prime}) - \theta (\Gamma)) + n^{\prime} \\
      &=
   2i\ \text{Im}\hspace{1pt}\left(\text{Tr}   \int_{M^{\prime} \rightarrow \Gamma} \frac{d \vec{k}}{2 \pi } \cdot \vec{\bm{A}} (\vec{k})\right) \\
   &\quad -i  n_{\text{occ}}
   +\frac{1}{2\pi}\sum_{i=1}^{n_{\text{occ}}} (\theta (M^{\prime}) - \theta (\Gamma))+  n^{\prime},
  \end{split}
\end{equation}
where $n^{\prime}$ denotes the integer.
Since the high-symmetry point $M^{\prime}$ is identical to $M$, we obtain Eq.~(\ref{eq:13}).

\section{Evaluation of the $\mathbb{Z}_4$ Berry phase in the presence of exceptional points \label{sec:eval-mathbbz_4-berry}}
\begin{figure*}[t]
 \centering
  \centering
\includegraphics[trim={0cm 0cm 0cm 0cm},width =0.9\hsize]{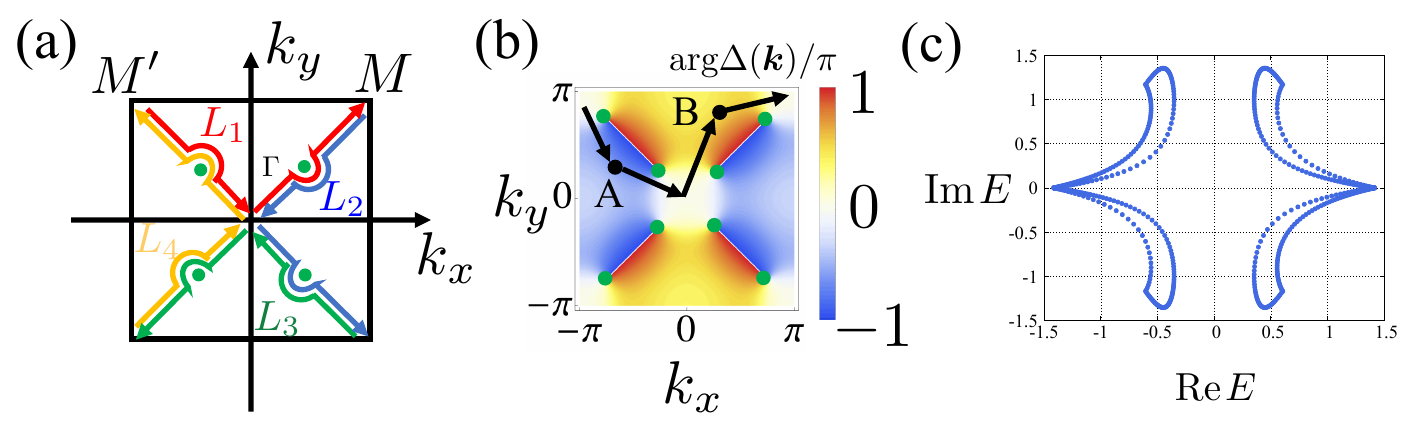}
 \caption{(Color online)
 (a) Sketch of the Brillouin zone and the integration path to avoid exceptional points.
 Green dots represent exceptional points.
 (b) Color plot of the map of $\text{arg}\Delta(\bm{k})/\pi$ for the model in Eq.~(\ref{eq:17}).
 Green dots are exceptional points.
 These points emerge at edges of the brunch cuts, i.e., $\Delta (\bm{k})=0$.
 Black dots A and B denote the momentum points $(k_x,k_y)= (-\pi/2,\pi/4)$, and $(\pi/4,\pi/2)$, respectively.
 (c) Energy spectra under the periodic boundary condition along the black arrows in (b).
 }
 \label{fig:Z4_EP}
\end{figure*}

We show the evaluation of the $\mathbb{Z}_4$ Berry phase in the presence of gapless points on the line for $k_x = \pm k_y$
except the high-symmetry points $\Gamma$ and $M$.
Specifically, we consider the evaluation of the complex Berry phase in Eq.~(\ref{eq:7}) along the path in Fig.~\ref{fig:Z4_EP} (a).
We show that,
even in the presence of exceptional points,
we can evaluate the $\mathbb{Z}_4$ Berry phase by the same approach in the above derivation
as far as there exists the real line gap between energy spectra along the path in Fig.~\ref{fig:Z4_EP} (a).

Let us see the representative example of the two-dimensional toy model with generalized four-fold rotational symmetry.
Specifically, we consider the Hamiltonian,
\begin{equation}
\label{eq:41}
 {\small
  h(\bm{k}) =
  \begin{pmatrix}
   \cos k_x + \cos k_y & (1-i)-i \cos k_x + i \cos k_y \\
    (-1-i)+i \cos k_x - i \cos k_y  & -\cos k_x - \cos k_y
  \end{pmatrix}
  }
\end{equation}
The above model respects generalized four-fold rotational symmetry,
\begin{equation}
\label{eq:42}
  U_4 h(\bm{k}) U_4^{-1} =h^{\dagger}(R_4 \bm{k}),
\end{equation}
where
\begin{equation}
\label{eq:43}
  U_4 =
  \begin{pmatrix}
   i & 0 \\
   0 & -i
  \end{pmatrix},
\end{equation}
denotes the unitary matrix.

First, we see the emergence of exceptional points.
The explicit forms of eigenvalues $E_{\pm}(\bm{k})$ of $h(\bm{k})$ are
\begin{equation}
\label{eq:44}
  E_{\pm}(\bm{k})=\pm
  \sqrt{
  2i \cos k_x
  -2i \cos k_y
  +\cos (2k_x)
  +\cos (2k_y)
  }.
\end{equation}
To show the emergence of exceptional points,
we show the numerical result of the discriminant $\Delta (\bm{k})$ of the characteristic polynomial $\det (h(\bm{k}) - E 1_{N \times N})$
[see Fig.~\ref{fig:Z4_EP} (b)].
In Fig.~\ref{fig:Z4_EP} (b),
exceptional points emerge at edges of the branch cuts, i.e., $\Delta (\bm{k})=0$.
From these numerical results,
at the momentum points $(k_x,k_y)= (\pi/4,\pm \pi /4)$, $(- \pi/4,\pm \pi /4)$, $( 3\pi/4,\pm 3\pi /4)$, and $(- 3\pi/4,\pm 3\pi /4)$,
there exist exceptional points ($E_{+}(\bm{k}) = E_{-}(\bm{k})$).

Second,
we show the presence of the real line gap and characterize the real line gap topology.
Figure~\ref{fig:Z4_EP} (c) displays the energy spectra under the periodic boundary condition along the black arrows in Fig.~\ref{fig:Z4_EP} (b).
We consider this path to avoid the exceptional points.
From this numerical result, we observe the presence of the real line gap,
which enables us to characterize the topology of the real line gap.
At the high-symmetry points $\Gamma$ and $M$,
the right eigenvectors of $E_{-}(\bm{k})$ are
\begin{align}
 \vec{\psi}_{-} (\Gamma)
 &=
 \left(
 \frac{(1-i)(-2+\sqrt{2})}{2\sqrt{4-2\sqrt{2}}},
 \frac{1}{2\sqrt{4-2\sqrt{2}}}
 \right), \label{eq:45}   \\
 \vec{\psi}_{-} (M)
 &=
 \left(
 \frac{(1+i)(2+\sqrt{2})}{2\sqrt{2+2\sqrt{2}}},
 -\frac{i}{\sqrt{2+2\sqrt{2}}}
 \right). \label{eq:46}
\end{align}
Thus, the explicit forms of $\theta_{-}(\bm{k})$ at the high-symmetry points $\Gamma$ and $M$ [see Eq.~(\ref{eq:14})] are
\begin{equation}
\label{eq:47}
  \theta_{-}(\Gamma)
  =
  \frac{1}{4},
  \quad
  \theta_{-}(M)
  =-\frac{1}{4}.
\end{equation}
These results indicate the emergence of the nontrivial phase of the $\mathbb{Z}_4$ Berry phase, i.e., $\text{Re}\hspace{1pt}\nu = 1/2$.

\section{Emergence of four topological phases characterized by the $\mathbb{Z}_4$ Berry phase \label{sec:mathbbz_4-topol-phas}}

\begin{figure*}[t]
 \centering
  \centering
\includegraphics[trim={0cm 0cm 0cm 0cm},width =\hsize]{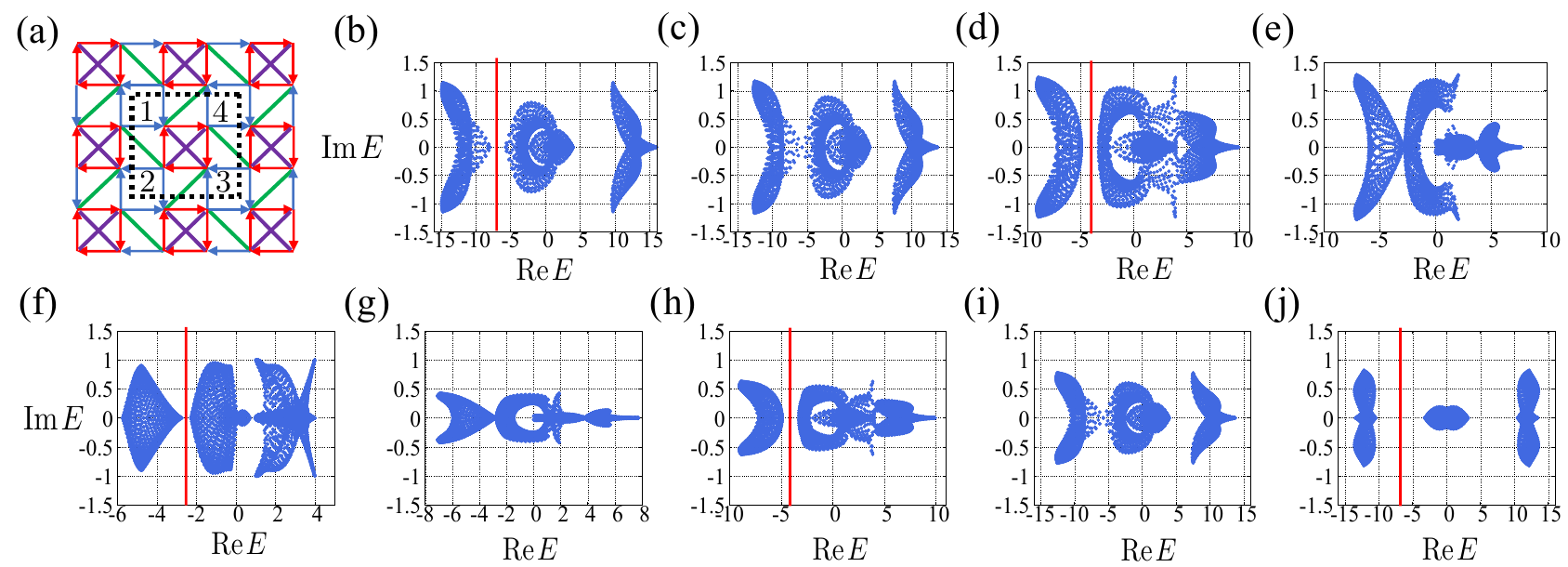}
 \caption{(Color online)
 (a) Schematic figure of the two-dimensional toy model with nonreciprocal hoppings in Eq.~(\ref{eq:48}).
 The unit cell is enclosed by the black dashed box.
 (b), (c), (d), (e), (f), (g), (h), (i), and (j) show
 energy spectra under the periodic boundary condition for
 several parameters.
 Panels (b), (c), (d), (e), (f), (g), (h), (i), and (j) display
 the energy spectra for
 $(t_1,t_2,t_3,t_4,\gamma_1,\gamma_2)=(1,-6,5/2,3/2,1/2,1/2)$,
 $(1,-\sqrt{17+4\sqrt{3}},5/2,3/2,1/2,1/2)$,
 $(1,-3,5/2,3/2,1/2,1/2)$,
 $(1,-\sqrt{2+\sqrt{3}},5/2,3/2,1/2,1/2)$,
 $(1,0,5/2,3/2,1/2,1/2)$,
 $(1,\sqrt{2+\sqrt{3}},5/2,3/2,1/2,1/2)$,
 $(1,3,5/2,3/2,1/2,1/2)$,
 $(1,\sqrt{17+4\sqrt{3}},5/2,3/2,1/2,1/2)$,
 and
 $(1,6,5/2,3/2,1/2,1/2)$,
 respectively.
 Red lines represent the real line gap where we focus.
 }
 \label{fig:Z4_FourPhases}
\end{figure*}

Let us see the emergence of the four topological phases characterized by the $\mathbb{Z}_4$ Berry phase.
Specifically, we consider the two-dimensional toy model in Fig.~\ref{fig:Z4_FourPhases} (a).
The red and blue lines denote the nonreciprocal hoppings.
Specifically, the hoppings along the red and blue arrows are $i(t_1 + \gamma_1)$ and $t_2 + \gamma_2$, respectively.
The hoppings opposite to the red and blue arrows are $-i(t_1 - \gamma_1)$ and $t_2 - \gamma_2$, respectively.
The purple and green lines denote the reciprocal hoppings.
Specifically, the hoppings along the purple (green) lines are $t_3$ ($t_4$).
Here, the explicit form of the Hamiltonian is
\begin{equation}
\label{eq:48}
  h(\bm{k})
  =
  \begin{pmatrix}
   h_{11}(\bm{k}) & h_{12} (\bm{k}) \\
   h_{21}(\bm{k}) & h_{22} (\bm{k})
  \end{pmatrix},
\end{equation}
with
\begin{equation}
 \label{eq:18}
  h_{11} (\bm{k}) =
  \begin{pmatrix}
   0 & i(t_1 - \gamma_1) + (t_2 - \gamma_2)e^{-i k_y} \\
   -i (t_1 + \gamma_1)+(t_2 + \gamma_2) e^{ik_y} & 0
  \end{pmatrix},
\end{equation}
\begin{equation}
 \label{eq:19}
  h_{12} (\bm{k}) =
  \begin{pmatrix}
    t_3 + t_4 e^{-i k_x} & - i (t_1 - \gamma_1) + (t_2 - \gamma_2)e^{i k_x} \\
   i (t_1 + \gamma_1) + (t_2 + \gamma_2)e^{i k_x} & t_3 + t_4 e^{-i k_y}
  \end{pmatrix},
\end{equation}
\begin{equation}
 \label{eq:20}
  h_{21} (\bm{k}) =
  \begin{pmatrix}
   t_3 + t_4 e^{ik_x} & - i (t_1 - \gamma_1) + (t_2 - \gamma_2) e^{-i k_x} \\
   i (t_1 + \gamma_1) + (t_2 + \gamma_2) e^{-ik_x} & t_3 + t_4 e^{ik_y}
  \end{pmatrix},
\end{equation}
and
\begin{equation}
\label{eq:25}
  h_{22} (\bm{k}) =
  \begin{pmatrix}
   0 & i (t_1- \gamma_1) + (t_2 - \gamma_2)e^{ik_y} \\
   - i (t_1 + \gamma_1) + (t_2 + \gamma_2) e^{-ik_y} & 0
  \end{pmatrix}.
\end{equation}

The above model respects generalized four-fold rotational symmetry,
\begin{equation}
\label{eq:49}
  U_4 h (\bm{k}) U_4^{-1} = h^{\dagger} (R_4 \bm{k}),
\end{equation}
with
\begin{equation}
\label{eq:50}
  U_4
  =
  \begin{pmatrix}
   0 & 0 & 0 & 1 \\
   1 & 0 & 0 & 0 \\
   0 & 1 & 0 & 0 \\
   0 & 0 & 1 & 0
  \end{pmatrix}.
\end{equation}

We observe the presence of the real line gap topology for $(t_1,t_3,t_4,\gamma_1,\gamma_2)=(1,5/2,3/2,1/2,1/2)$.
To discuss the presence of real line gap, we plot energy spectra on the complex plane (see Fig.~\ref{fig:Z4_FourPhases}).
For $t_2 \neq \pm \sqrt{2+\sqrt{3}}$ and $\pm \sqrt{17+4\sqrt{3}}$,
there exists the real line gap represented by the red lines in Figs.~\ref{fig:Z4_FourPhases} (a), (c), (e), (g), and (i).
Here, we note that the gapless points at $\Gamma$ and $M$ can be obtained from the computation of the discriminant of the characteristic polynomial.
By characterizing the real line gap topology on the red line in Fig.~\ref{fig:Z4_FourPhases},
we obtain the $\mathbb{Z}_4$ Berry phase
\begin{equation}
\label{eq:51}
  \text{Re}\hspace{1pt}\nu
  =
  \begin{cases}
   \frac{1}{2} & \text{for }t_2 <  -\sqrt{17+4\sqrt{3}}  \\
   \frac{1}{4} & \text{for }   -\sqrt{17+4\sqrt{3}} <t_2< -\sqrt{2+\sqrt{3}}  \\
   0 & \text{for } -\sqrt{2+\sqrt{3}} < t_2 <\sqrt{2+\sqrt{3}}\\
   \frac{3}{4} & \text{for } \sqrt{2+\sqrt{3}}<t_2 < \sqrt{17+4\sqrt{3}} \\
   \frac{1}{2} & \text{for } \sqrt{17+4\sqrt{3}} < t_2
  \end{cases}.
\end{equation}
Here, we have used the fact that
the $\mathbb{Z}_4$ Berry phase does not change without the gap closing at $\Gamma$ and $M$.
These results show the emergence of four topological phases.
We note that the physical picture of the adiabatic connection and the resulting the bulk-edge correspondence
are not straightforwardly obtained in this model.

\section{Corner states characterized by $(\text{Re}\hspace{1pt} \nu , w)=(1/2,1)$ and $(0,-1)$ \label{sec:corn-stat-char}}

\begin{figure}[b]
 \centering
  \centering
\includegraphics[trim={0cm 0cm 0cm 0cm},width =\hsize]{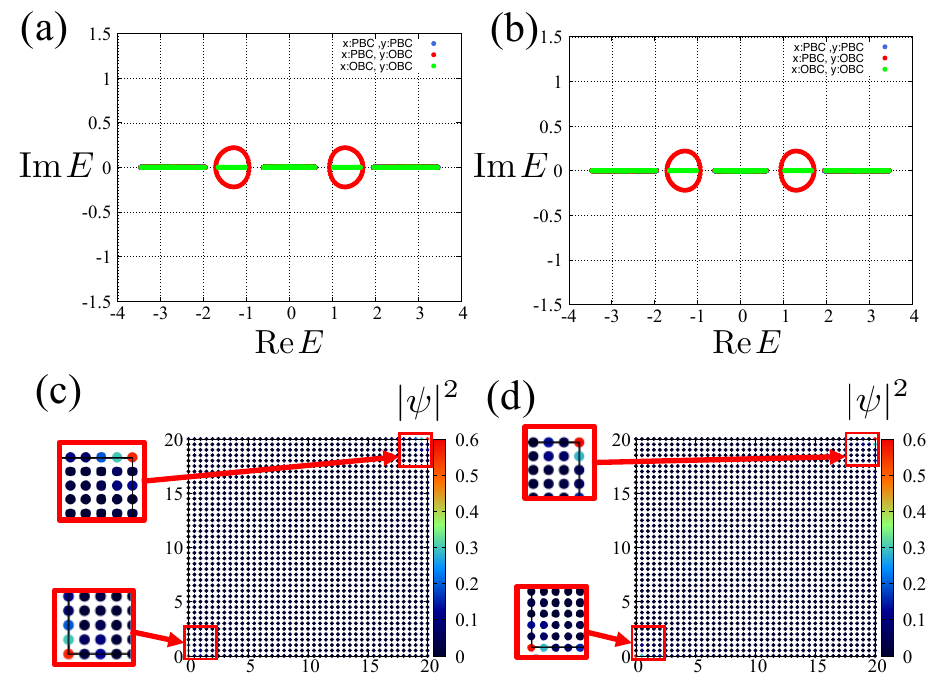}
 \caption{(Color online)
 (a) and (b) show energy spectra on the complex plane for $(\delta t , \delta \gamma) = (-0.7,-0.5)$ and $(0.7,0.5)$, respectively.
 Here, the blue dots show the energy spectra under the periodic boundary condition along both $x$ and $y$ directions.
 The red (green) dots show the energy spectra under the periodic (open) boundary along $x$ direction and the open boundary condition along $y$ direction.
 (c) and (d) show the distributions of the eigenstates for $(\delta t, \delta \gamma, E) = (-0.7,-0.5,-1.45087)$ and $(0.7,0.5,-1.31982)$, respectively.
 }
 \label{fig:Z4_Other}
\end{figure}

In this section, we see the emergence of the corner states for $(\text{Re}\hspace{1pt} \nu , w)=(1/2,1)$ and $(0,-1)$.
Figures~\ref{fig:Z4_Other} (a) and (b) display the energy spectra for $(\delta t , \delta \gamma) = (-0.7,-0.5)$ and $(0.7,0.5)$, respectively.
The blue dots show the energy spectra under the periodic boundary condition along both $x$ and $y$ directions.
In addition, the red (green) dots show the energy spectra under the periodic (open) boundary along $x$ direction and the open boundary condition along $y$ direction.
From the result in Fig.~\ref{fig:SSH2D_PBC} (f), the $\mathbb{Z}_4$ Berry phase for $(\delta t , \delta \gamma) = (-0.7,-0.5)$ $[(0.7,0.5)]$ takes the value $\text{Re} \hspace{1pt} \nu = 1/2$ $(0)$.
By the same approach of the adiabatic connection in Sec.~\ref{sec:edge-spectr-cylind},
for $(\delta t , \delta \gamma) = (-0.7,-0.5)$ $[(0.7,0.5)]$,
edge states emerge at the top and left [bottom and right] side.
Furthermore, there exists the point gap of the edge spectra characterized by the winding number $w=1$ ($-1$) for $(\delta t , \delta \gamma) = (-0.7,-0.5)$ $[(0.7,0.5)]$.
Therefore, from the presence of the nontrivial point gap topology of edge spectra,
we see the emergence of hybrid higher-order skin topological modes localized at the right top and left bottom corners [see Figs.~\ref{fig:Z4_Other} (c) and (d)].
Here, we note that the corner states for $(\delta t , \delta \gamma) = (-0.7,-0.5)$ $[(0.7,0.5)]$
arise from the edge states localized at the top and left [bottom and right] side.
By combining these results,
we find that the pair of topological invariants for $(\text{Re}\hspace{1pt} \nu , w)=(1/2,1)$ and $(0,-1)$
predicts the emergence of hybrid higher-order skin topological modes localized at the right top and left bottom corners.


\begin{thebibliography}{100}
\bibitem{footnote1} We note that the bulk-boundary correspondence for the $\mathbb{Z}_4$ Berry phase is not always straightforward unless simple adiabatic limits exist.
\end{thebibliography}
\end{document}